\documentclass{aa}  
\usepackage{graphicx}
\usepackage{txfonts}
\begin{document}
   \title{Energetics of solar coronal mass ejections}

   \author{Prasad Subramanian
          \inst{1}
          \and
          Angelos Vourlidas\inst{2}
          }

   \offprints{Prasad Subramanian}

   \institute{Indian Institute of Astrophysics, Koramangala, Bangalore - 560034, India\\
              \email{psubrama@iiap.res.in}
         \and
             Code 7663, Naval Research Laboratory, Washington, DC 20375, USA\\
             \email{vourlidas@nrl.navy.mil}}

   \date{}

 
  \abstract
   {}
   {We investigate whether solar coronal mass ejections are driven
mainly by coupling to the ambient solar wind or through the release of internal magnetic energy.}
   {We examine the energetics of 39 flux-rope like coronal mass
  ejections (CMEs) from the Sun using data in the distance range $\sim$ 2--20 $R_{\odot}$ 
from the Large Angle
  Spectroscopic Coronograph (LASCO) aboard the Solar and Heliospheric
  Observatory (SOHO).  This comprises a complete sample of the best
  examples of flux-rope CMEs observed by LASCO in 1996-2001.}
   {We find
  that 69\% of the CMEs in our sample experience a clearly identifiable driving power in
  the LASCO field of view.  For the CMEs that are driven, we
  examine if they might be deriving most of their driving power by
  coupling to the solar wind. We do not find conclusive evidence in
  favor of this hypothesis. On the other hand, we find that their internal
  magnetic energy is a viable source of the required
  driving power. We have estimated upper and lower limits on the power that can
possibly be provided by the internal magnetic field of a CME. We find that, on average,
the lower limit to the available magnetic power is around 74\% of what is required to
drive the CMEs, while the upper limit can be as much as an order of magnitude larger.}
   {}

   \keywords{}

   \maketitle
%

\section{Introduction}

The basic energetics of coronal mass ejections (CMEs) from the Sun is
a subject of intense research.  The amount of energy required to
disrupt initially closed magnetic field lines and to lift and accelerate
CMEs against the gravitational field of the Sun are key ingredients of CME initiation models
(e.g., Amari et al. 2000; Antiochos, DeVore \& Klimchuk 1999; Forbes
2000). While the energetics of CMEs in the lower corona is poorly
understood, the energetics of CMEs beyond $\sim$ 2R$_{\odot}$ is
somewhat better understood (Vourlidas et al. 2000; Vourlidas et al.
2002; Lewis \& Simnett 2002). Since the advent of the excellent dataset of CMEs
provided by the Large Angle Spectroscopic Coronograph (LASCO,
Brueckner et al. 1995) aboard the Solar and Heliospheric Observatory
(SOHO, Domingo et al. 1995), there have been only a few papers that have
examined the energetics of several CMEs.
Vourlidas et al. (2000) (Paper 1 from now on) studied the evolution of
the potential, kinetic, and magnetic energies of 11 flux-rope CMEs in
an attempt to understand the driving mechanism for such CMEs beyond
$\sim$ 2R$_{\odot}$. They surmised that the energy contained in the
magnetic fields advected by the CMEs could be responsible for
propelling them. Some recent studies of the initiation of flux-rope CMEs (Amari et al.
2000) suggest that $\sim$ 55\% of the available magnetic free energy
could be available for propagating the CME through the corona.

On the other hand, Lewis \& Simnett (2002) used an ingeneous method to
study the weighted average profile of all the CMEs in the LASCO C2 and
C3 fields of view from March 1999 to March 2000 to
investigate similar questions. They found that the mechanical (i.e.,
kinetic + potential) energy of a typical CME in this period increased
with time at a remarkably constant linear rate as it propagated
through the LASCO C2 and C3 fields of view. Based on this constant
rate of input power to a typical CME, they concluded that CMEs are
likely to be powered by momentum coupling with the solar wind, which
is an effectively infinite energy reservoir for most CMEs. It may
  be noted that they did not measure individual CMEs to arrive at this
  conclusion, nor did they present adequate calculations to support it. It
  is therefore an aggregate statement and, as we will see later, an
  incorrect one. In contrast, our method, which is outlined in \S~2,
  involves detailed measurements of each CME in our sample.
Manoharan (2006) has studied the evolution of 30 large Earth-directed
CMEs by combining data from LASCO with that from the Ooty Radio
Telescope (ORT).  His dataset spans distances from $\sim$ 2
$R_{\odot}$--1 AU. He notes that the average CME in his sample arrives
at the Earth around 13 hours sooner than a typical parcel of solar
wind would, and thereby concludes that CMEs are not simply dragged
along by the solar wind; they have to be driven by the expenditure of
some kind of internal energy. However, the CMEs in his sample slow
down significantly at distances $>$ 80 $R_{\odot}$.  This suggests
that the solar wind might be influencing CME propagation significantly
for $R > 80 R_{\odot}$.

In this work, we concentrate on flux-rope (FR) CMEs because (i)
flux-ropes are commonly invoked by several current theoretical and
numerical models of CMEs (e.g., Chen 1996; Kumar \& Rust 1996; Gibson
\& Low 1998; Birn, Forbes \& Schindler 2003; Kliem \& T\"or\"ok 2006) and (ii) their physical
parameters can be derived by in-situ observations (e.g., Burlaga 1988;
Lepping et al 1990; Hu and Sonerup 1998; Mulligan and Russel 2001;
Lynch et al. 2003; Lepping et al 2003).  Generally, LASCO observes
many events sufficiently structured to be characterized as FR CMEs
under some viewing assumptions (e.g., Cremades and Bothmer 2004). In
Paper 1 and here, we have adopted a much stricter definition for a
FR CME; namely, the event must exhibit a clear circular structure with
visible striations in its core. In other words, the CME must closely resemble the cross section of a theoretical flux-rope (also see \S~3.2). Based on this
criterion, we study the evolution of potential and kinetic energies of
39 individual FR CMEs between 1997 and 2001.  This comprises a
complete sample of the best examples of FR CMEs observed by LASCO in
1996-2001 (out of about 4000 events).  In doing so, we obtain better
statistics than Paper 1 and include a wider variety of events through
the rising phase and maximum of cycle 23.  We find that the mechanical
energy (i.e., kinetic + potential energy) of 69\% of the events
increases linearly with time. This implies that these events are
clearly ``driven'' by the release of some sort of energy.  Based on
our examination of these individual events, we investigate if the CMEs
could be powered by coupling to the solar wind. We also examine whether
the release of the internal magnetic energy of a CME can
account for its driving power.

\section{Data analysis}
\subsection{Mass images}
We have compiled a complete list of
all CMEs that appear like flux ropes in the LASCO data between
February 1997 and March 2001 and selected the best cases based on their
morphological appearance in coronograph images for this
study. The Thomson scattering process by which free electrons in the
CME scatter photospheric light and give rise to these intensity images
has a rather sharp dependence on the scattering angle. Those CMEs that
retain their overall morphology in LASCO images 
are therefore probably ones that
remain in the plane of the sky throughout these fields of view
(Cremades and Bothmer 2004; also see \S~3.2). 
Since the calculations of CME mass (see
Paper 1) assumed that the CME is in the plane of the sky, this lends
credence to our estimates of CME mass and velocity.

We now briefly
describe the procedure we followed in order to obtain the evolution of
CME energy from a time sequence of these intensity images.  The
intensity of Thomson-scattered light depends directly on the column
density of coronal electrons off of which the scattering takes place. By
backtracking through the Thomson scattering calculations, we are thus
able to construct mass images from the observed intensity images.
Each pixel of the mass image gives the surface density (g cm$^{-2}$)
of coronal electrons. By subtracting a suitable pre-event (or, in some
cases, post-event), mass image from the image containing the CME, we
obtain an image that gives the excess mass (over the background
corona) carried by the CME. We circumscribe the extent of the
flux-rope structure within the CME as evident in the image and get its
total mass by simply summing the masses of all the pixels comprising
the CME. It is also straightforward to obtain the center of mass for
the flux-rope structure of the CME from such a mass image, since we
know the mass contained in each pixel and its spatial co-ordinates. A
time sequence of these mass images gives the evolution of CME mass and
the velocity of the center of mass. The time evolution of kinetic and
potential energies of the CME are calculated from these quantities.
This part of the data analysis procedure is similar to what is used in Paper 1,
and we refer the reader there for further details.

\begin{figure*}
\centering
\includegraphics[width=12cm]{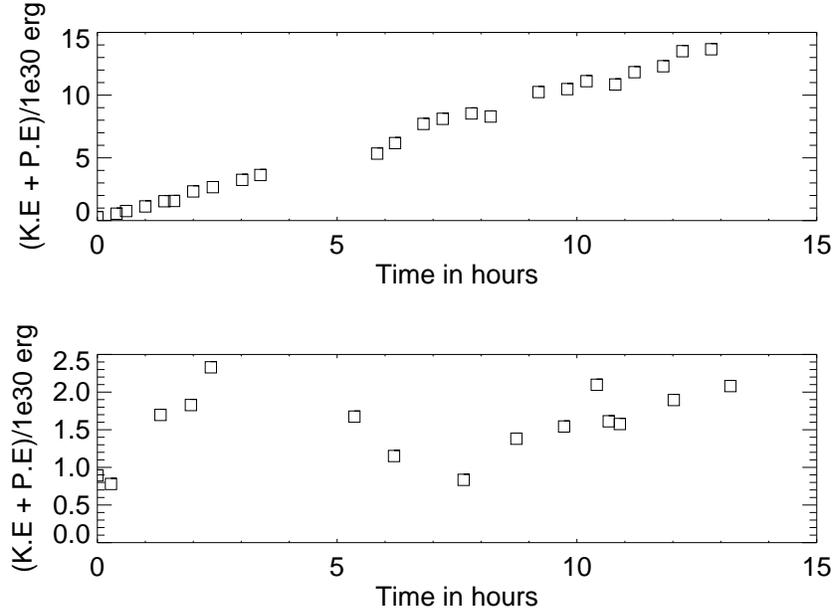}
\caption{The mechanical
  (i.e., kinetic + potential) energy for two representative CMEs
  plotted as a function of time from initiation. The mechanical energy
  for the CME on 2000/03/22 (upper panel) increases linearly with time,
  implying that there is a constant driving power on the CME as it
  propagates outwards. Such CMEs are included in category A (Table 1).
  The mechanical energy for the CME on 1998/08/13 (lower panel) shows
  no such trend. Such CMEs are included in category B (Table 2).}
\label{}
\end{figure*}

\subsection{Driving power}
Having obtained the time evolution of the kinetic and potential
energies of a CME, we add them together to obtain the time evolution
of its mechanical (i.e., kinetic + potential) energy.  We find that
for 27 CMEs, the mechanical energy rises linearly with time (category
A, Table 1), whereas 12 CMEs show no such trend (category B, Table 2).
In other words, 27 out of 39 CMEs (69\%) belong to category A, whereas
the remaining 12 (31\%) CMEs belong to category B.  The upper panel of
figure 1 shows an example of a CME in category A (Table 1), where the
linear rise of mechanical energy with time is clearly evident. The
lower panel of figure 1 shows an example of a CME in category B (Table
2). 

\begin{table*}
\caption{Category {\bf A}: CMEs for which Mechanical Energy
increases linearly with time}             
\label{table:1}      
\centering 
\renewcommand{\footnoterule}{}  
\begin{tabular}{c c c c c c c}        
\hline\hline                 
Date & Time & PA & Speed & At Radius & Mass & Eruptive Prominence\\
\hline 
 &  & $\degr$ & (km/s) & (R$_{\odot}$) & ($\times 10^{15}$ g) & \\ 
\hline                        
97/11/01  & 20:11 & 271 & 275 & 20     & 1   & Y\\
97/11/16  & 23:27$^{\mathrm{a}}$ & 85 & 595 & 20.5   & 5   & N\\ 
98/02/04  & 17:02 & 289 & 425 & 19.5   & 5   & N\\
98/02/24  & 07:28 & 90 & 500 & 19     & 1   & N\\
98/05/07  & 11:05 & 270 & 450 & 21     & 10  & N\\
98/06/02  & 08:08 & 245 & 600 & 14.5   & 10  & Y\\
99/07/02  & 17:30 & 39 & 220 & 16.5   & 5   & Maybe\\
99/08/02  & 22:26 & 271 & 380 & 24     & 2.5 & Y\\
00/03/22  & 04:06 & 323 & 350 & 14     & 5   & Maybe\\
00/05/05  & 07:26 & 338 & 260 & 9      & 1   & N\\
00/05/29  & 04:30 & 278 & 178 & 10     & 1.5 & Maybe\\
00/06/06  & 04:54 & 359 & 400 & 15     & 4   & Y\\
00/06/08  & 17:07 & 59 & 310 & 10.5   & 2   & N\\
00/07/23  & 17:30 & 14 & 400 & 9      & 1   & N\\
00/08/02  & 17:54 & 46 & 700 & 20     & 7   & Y\\
00/08/03  & 08:30 & 302 & 620 & 18     & 6   & Y\\
00/09/27  & 00:50 & 327 & 455 & 15     & 1   & N\\
00/10/26  & 00:50 & 99 & 200 & 12.5   & 2   & Maybe\\
00/11/12  & 09:06 & 329 & 282 & 15     & 2   & N\\
00/11/14  & 16:06 & 258 & 500 & 20     & 2   & N\\
00/11/17  & 04:06 & 75 & 450 & 18 & 3 & Y\\
00/11/17  & 06:30 & 188 & 500 & 18 & 3 & Y\\
01/01/07  & 04:06 & 298 & 550 & 17     & 3   & Y\\
01/01/19  & 17:06 & 78 & 900 & 18     & 3   & Maybe\\
01/02/10  & 23:06$^{\mathrm{a}}$ & 229 & 900 & 23     & 4   & Y\\
01/03/01  & 04:06 & 292 & 400 & 20     & 1   & Maybe\\
01/03/23  & 12:06 & 284 & 400 & 15     & 7   & N \\
\hline                                   
\end{tabular}
\begin{list}{}{}
\item[$^{\mathrm{a}}$] The time refers to the previous day.
\item[] \textsl{Column 1\/}: Date on which a given CME
  occurred; \textsl{Column 2\/}: Start time in the C2 field of view;
  \textsl{Column 3\/}: central position angle of the CME (CCW from
  solar north); \textsl{Column 4\/}: Speed of the CME at the radius
  quoted in column 5; \textsl{Column 5\/}: This is the farthest radius
  until which we have been able to track the CME; \textsl{Column 6\/}:
  Mass of the CME at the radius quoted column 5. For instance, the CME
  on 97/11/01 has a speed of 275 km/s and a mass of $10^{15}$ g at 20
  R$_{\odot}$; \textsl{Column 7\/}: Denotes whether or not the CME was
  associated with a prominence eruption (see~\S~3.4); `Y' denotes that
  the CME was associated with a prominence eruption, `N' denotes the
  converse and `Maybe' denotes a situation where we are not certain
  that a prominence eruption was associated with the CME.
\end{list}
\end{table*}

\begin{table*}
\caption{Category {\bf B}: CMEs for which mechanical energy
remains constant with time}             
\label{table:2}      
\centering 
\renewcommand{\footnoterule}{}  
\begin{tabular}{c c c c c c c}        
Date & Time & PA & Speed & At Radius & Mass &Eruptive
  Prominence\\
\hline
 &  & ($\degr$) & (km/s) & (R$_{\odot}$) & ($\times 10^{15}$ g) & \\ 
\hline\hline                 
97/02/23 &  02:55 & 82 & 910 & 15.5 & 1   & Y\\
97/04/13 &  16:12 & 269 & 510 & 24   & 0.8 & Y\\
97/04/30 &  04:50 & 84 & 330 & 18.5 & 0.7 & N\\
97/08/13 &  08:26 & 273 & 350 & 20   & 1   & N\\
97/10/19 &  04:42 & 92 & 260 & 11   & 1   & Y\\
97/10/30$^{\mathrm{a}}$ &  18:21 & 88 & 225 & 17.5 & 1   & N\\
97/10/31 &  09:30 & 262 & 410 & 23   & 1   & N\\
99/05/23 &  07:40 & 288 & 600 & 30   & 1   & N\\
99/07/04 &  21:54$^{\mathrm{a}}$ & 89 & 181 & 16   & 2   & Maybe\\
00/11/04 &  01:50 & 213 & 794 & 29   & 3   & Y\\
01/01/19 & 12:06 & 74 & 403& 17  & 1   & N\\
01/03/22 &  05:26 & 255 & 377 & 14.5 & 2   & Y\\
\hline                                   
\end{tabular}
\begin{list}{}{}
\item[$^{\mathrm{a}}$] The time refers to the previous day.
\item[] Columns same as Table 1.
\end{list}
\end{table*}
 
For the CMEs in category A (Table 1), we fit a straight line to
the plot of mechanical energy vs. time. The slope of this straight
line gives the driving power. As pointed out in paper 1 (also see
Vourlidas 2004, Lugaz et al. 2005) the mass of a given CME can be
underestimated by at most a factor of 2. Furthermore, this would be a
systematic error in the mass estimate for a given CME. It does not
affect the {\em slope} of the mechanical energy vs. time curve for a
given CME.  The errors $\sigma_{D}$ on the values of the driving power
thus arise only from the errors in determining the slope of the
straight line fit. Column 2 of Table 3 gives the driving power $P_{D}$
determined in this manner and column 3 of Table 3 gives the
associated error $\sigma_{D}$ for each CME in category A. Both these
quantities are expressed in units of $10^{30}$ erg/hr.

\subsection{Estimate of magnetic power}
The driving power could be provided by the release of the internal magnetic
energy of the FR CMEs.  In order to estimate the power that can
possibly be released by magnetic fields advected with an expanding
CME, we need to know the magnetic field advected with the CME.

\subsubsection{Direct estimate of magnetic fields carried by CMEs}
Measurements of the coronal magnetic field (much less so for the
magnetic field entrained by CMEs) are few and far between. 
Using radio measurements of what is presumably synchrotron
emission from electrons populating the CME structure, 
Bastian et al. (2001) have estimated the
magnetic field in a CME on 1998 April 20 to be $\sim$ 0.1 -- 1 G.
We adopt the value of 0.1G as a working figure for our purposes.

The magnetic energy contained in the CME can be written as
\begin{equation}
\widetilde{E_{M}} = \frac{B^{2}}{8\,\pi}\,l\,A \, ,
\end{equation}
where $B$ is the magnetic field, $A$ the cross-sectional area of the CME, and $l$ its length perpendicular to the
plane of the sky. We measure $A$ directly for each CME (in each image), and we take $l$ to be equal to the heliocentric
distance of the CME center of mass (in each image). 
The assumption for $l$ implies a reasonable flux rope length of
one solar radius at the solar surface. The power ($\widetilde{P_{M}}$) that can possibly be released by the advected magnetic field is 
\begin{equation}
\frac{d}{dt}\,\widetilde{E_{M}} = \widetilde{P_{M}} = \frac{B^{2}}{8\,\pi}\,\frac{d}{dt}\,l\,A\, .
\end{equation}
Note that we have not accounted for the temporal variation of the magnetic field $B$ in computing $\widetilde{P_{M}}$.
We use a conservative value of 0.1 G for the magnetic field $B$ and fit a
straight line to the time evolution of $l\,A$ to get the values of
$\widetilde{P_{M}}$ shown in column 8 of Table 3. The associated error
$\widetilde{\sigma_{M}}$ quoted in column 9 of Table 3 arises only
from the error $\sigma_{l\,A}$ in the straight line fit to the time
evolution of $l\,A$. The quantity $\widetilde{\sigma_{M}}$ is defined
as
\begin{equation}
\widetilde{\sigma_{M}} = \frac{B^{2}}{8\,\pi}\,\sigma_{l\,A} \, .
\end{equation}
The quantities $\widetilde{P_{M}}$ and $\widetilde{\sigma_{M}}$ are
expressed in units of $10^{30}$ erg/hr in table 3. Since we do not account for the possible decrease in the advected magnetic field as the CME propagates outwards, $\widetilde{P_{M}}$ is an {\em upper limit} on the
power that can possibly be provided by its dissipation.
\subsubsection{Magnetic flux carried by near-Earth magnetic clouds}
On the other hand, magnetic clouds observed by near-Earth spacecraft are thought to be
near-Earth manifestations of CMEs that are directed towards the Earth
(e.g., Webb et al. 2000; Berdichevsky et al. 2002; Manoharan et al.
2004). We envisage a scenario where 
some of the magnetic flux carried by a CME is expended in driving it; what is left when it arrives at the Earth 
is detected by in-situ measurements of the corresponding near-Earth magnetic cloud. 
We can compute the magnetic power by assuming 
that the CME carried the same amount
of magnetic flux near the Sun as what is observed in the near-Earth
magnetic cloud. Such a calculation will necessarily yield a {\em lower
limit} on the power that can be expended by the advected magnetic field
in driving the CME.

Since the CMEs in our sample propagate primarily along the plane of the sky,
they will not be detected as near-Earth magnetic clouds.
However, Lepping et al. (1997) estimate the average magnetic flux
carried by 30 well observed magnetic clouds to be $\overline{B.A} =
10.8 \times 10^{20}$ Mx, with a standard deviation error of
$\sigma_{BA} = 8 \times 10^{20}$ Mx. The value
of $\sigma_{BA}$ they quote is representative of the range of fluxes carried by
different magnetic clouds and not of the errors in individual
measurements. The actual fit error for $B$ and $A$ is approximately
$6-7\%$ (Lepping et al. 2003); it is insignificant in
comparison to the overall flux variation, $\sigma_{BA}$. 
If we assume that $\overline{B.A}$ is representative of the magnetic flux
carried by the CMEs in our sample, we can write the following
expression for the CME magnetic energy:

\begin{equation}
E_{M} = \frac{1}{8\,\pi}\,\frac{l}{A}\,(\overline{B.A})^{2} \, ,
\end{equation}
where $l$ and $A$ are the length and cross-sectional area of the flux
rope, respectively. We take $l$ equal to the heliocentric distance of
the CME as we did in (1). Consequently,
a lower limit on the power derived from the decrease in magnetic energy
as the flux rope expands outwards is given by

\begin{equation}
P_{M} = \frac{d}{dt}\,E_{M} = \frac{1}{8\,\pi}\,(\overline{B.A})^{2}
\,\frac{d}{dt}\,\frac{l}{A}\, . 
\end{equation}
We have information about the time derivative of the quantity $l/A$
for each of the CMEs in our sample. The values
of $P_{M}$ are quoted in column 4 of Table 3 in units of $10^{30}$
erg/hr for the CMEs in category A; i.e., the ones that show clear
evidence of a driving power. The quantity $\sigma_{M}$ quoted in
column 5 of Table 3 is the error in the value of the magnetic power,
expressed in units of $10^{30}$ erg/hr. The error $\sigma_{M}$ in the
value of the magnetic power arises from the error $\sigma_{BA}$ in the
average magnetic flux and the error $\sigma_{l/A}$ in fitting a
straight line to the time evolution of $l/A$. The error in the value
of $\overline{B.A}^{2}$ is related to $\sigma_{BA}$ by
$\sigma_{BA^{2}} = 2\,\,\overline{B.A}\,\,\sigma_{BA}$.  The value of
$\sigma_{M}$ is defined by
\begin{equation}
\sigma_{M} = P_M\sqrt{\biggl (\frac{\sigma_{BA^2}}{(\overline{B.A})^2}\biggr )^{2} + 
\biggl (\frac{\sigma_{l/A}}{\frac{d}{dt}\frac{l}{A}}\biggr )^2}\,
\end{equation}

\section{Results and Interpretation}

As mentioned earlier, the mechanical energies of
the 27 CMEs in category A (Table 1) increase linearly with time,
implying a constant driving power for these CMEs. The mechanical
energies for the 12 CMEs in category B (Table 2), on the other hand,
show no such trend.  Figure 1 shows an example from each category; the
upper panel shows an example of a CME for which the mechanical energy
increases linearly with time, implying a constant driving power, while
the lower panel shows an example where there is no evidence for a
linear increase of mechanical energy with time. 

\subsection{Source of driving power for CMEs from $\sim$ 2--20 $R_{\odot}$: solar wind or advected
magnetic field?} 
Based on the constancy of power required to drive a typical CME, Lewis
\& Simnett (2002) surmise that CMEs could be driven via momentum
coupling with the solar wind, which is an effectively infinite energy
reservoir for the CMEs. However, they did not
measure individual CMEs to arrive at this conclusion, but instead employed a
weighted average method that gave this result for a typical CME
between March 1999 and March 2000.

CMEs could be driven by the ambient solar wind via the hydromagnetic buoyancy force $F_{\rm solwind}$. 
We write the following expression for this force following eq~(22) of Yeh (1995):
\begin{equation}
F_{\rm solwind} = \pi\,Q^{2}\,(-\,\nabla\,p_{\infty})\, ,
\end{equation}
where
$\pi\,Q^{2}$ represents the cross-sectional area presented by the CME, and the term inside
the brackets is the gradient in the ambient pressure that drives the solar wind. Evidently,
if the driving force on a CME is predominantly due to coupling with the solar wind, it should
be proportional to its cross-sectional area.
We now take a closer look at the CMEs
in category A (Table 1). Figure 2 is a scatterplot of the mechanical driving force versus mean CME size
(measured in number of pixels) for these CMEs.
\begin{figure*}
\centering
\includegraphics[width=12cm]{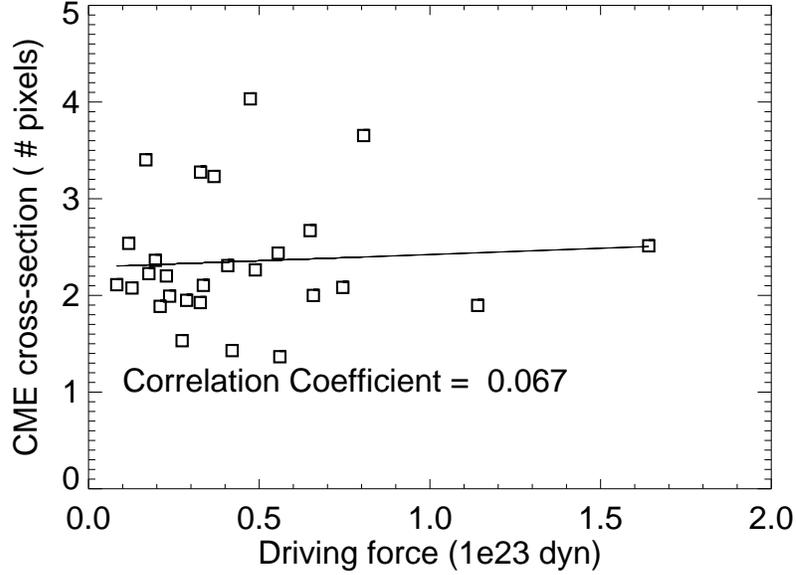}
\caption{The mean size (in number of pixels) for the CMEs in category
  A (Table 1) plotted as a function of their driving force. The low
  correlation coefficient suggests that there is no evidence to claim
  that larger CMEs have greater driving forces (see \S~3.1).}
\end{figure*} 
We calculate the driving force by dividing the driving power 
for a CME (\S~2.2) by the velocity of its center of mass.
The correlation between the driving
force and CME size is evidently poor, and there is little evidence to
suggest that larger CMEs experience a greater driving force. This casts doubt on
the hypothesis that the CMEs in category A (Table 1) (which are clearly ``driven'') 
are powered by coupling with the
ambient solar wind.

On the other hand, several researchers have suggested that a
combination of different kinds of Lorentz forces can drive the CME
outward (e.g., Chen 1996; Kumar \& Rust 1996). Most recently, Kliem \&
T\"or\"ok (2006), have investigated the interesting possibility of the
so-called torus instability being responsible for driving the CME.
This instability relies on the interplay between the Lorentz
self-force in the torus-like CME structure and the opposing Lorentz
force due to the ambient magnetic field.

We therefore turn our attention to the CME magnetic field to see if it
can act as a driver. In view of the considerable uncertainties in
determining coronal magnetic fields, we computed upper and lower
limits on the rate of energy released by the magnetic field advected
by each CME.  The procedures we adopted are explained in \S~2.3.1 and
\S~2.3.2.  We computed the magnetic powers only for the CMEs in
category A (Table 1), which are evidently driven.

\begin{table*}
\caption{Driving power and magnetic power for Category {\bf A}}             
\label{table:3}      
\centering 
\renewcommand{\footnoterule}{}  
\begin{tabular}{c c c c c c c c c c c}        
Date & $P_{D}$ & $\sigma_{D}$ & $P_{M}$ & $\sigma_{M}$ & 
$P_{M}$/$P_{D}$ & $\sigma_{P_{M}/P_{D}}$ & $\widetilde{P_{M}}$ & 
$\widetilde{\sigma_{M}}$ & $\widetilde{P_{M}}$/$P_{D}$ & 
$\sigma_{\widetilde{P_{M}}/P_{D}}$\\
\hline \hline
97/11/01  & 0.229 & 0.022 & 0.620 & 0.925 & 2.708 & 4.032 &  5.872 & 0.667 & 25.583 &  2.904\\
97/11/16  & 2.426 & 0.198 & 0.287 & 0.431 & 0.118 & 0.712 & 24.548 & 2.595 & 10.117 &  1.069\\
98/02/04  & 1.477 & 0.158 & 0.336 & 0.505 & 0.228 & 0.583 & 10.945 & 0.132 &  7.410 &  0.897\\
98/02/24  & 0.306 & 0.077 & 0.511 & 0.758 & 1.668 & 2.480 &  6.642 & 0.808 & 21.686 &  2.639\\
98/05/07  & 3.295 & 0.355 & 0.494 & 0.739 & 0.150 & 0.751 &  8.145 & 1.554 &  2.471 &  0.473\\
98/06/02  & 7.274 & 0.656 & 0.718 & 1.086 & 0.098 & 0.926 & 23.054 & 3.809 &  3.169 & 0.524\\
99/07/02  & 0.835 & 0.127 & 0.187 & 0.278 & 0.224 & 0.757 & 18.826 & 3.061 & 22.550 &  3.667\\
99/08/02  & 0.478 & 0.023 & 0.324 & 0.481 & 0.678 & 1.009 &  7.200 & 0.682 & 15.053 &  1.425\\
00/03/22  & 1.060 & 0.017 & 0.312 & 0.463 & 0.295 & 0.441 &  3.147 & 0.409 &  2.975 &  0.387\\
00/05/05  & 0.358 & 0.038 & 1.120 & 1.665 & 3.122 & 4.646 &  1.324 & 0.160 &  3.693 &  0.447\\
00/05/29  & 0.488 & 0.052 & 0.528 & 0.783 & 1.082 & 1.608 &  2.311 & 0.274 &  4.733 &  0.562\\
00/06/06  & 1.153 & 0.031 & 0.659 & 0.979 & 0.572 & 0.851 &  4.750 & 0.622 &  4.121 &  0.539\\
00/06/08  & 0.705 & 0.095 & 0.840 & 1.269 & 1.190 & 1.802 &  9.941 & 1.364 & 14.088 &  1.933\\
00/07/23  & 0.747 & 0.208 & 0.740 & 1.106 & 0.989 & 1.505 &  3.566 & 0.581 &  4.770 &  0.780\\
00/08/02  & 3.557 & 0.099 & 0.562 & 0.843 & 0.158 & 0.295 & 28.542 & 3.970 &  8.025 & 1.115\\
00/08/03  & 3.789 & 0.200 & 0.839 & 1.271 & 0.221 & 0.411 & 30.443 & 4.224 &  8.035 & 1.115\\
00/09/27  & 0.805 & 0.100 & 0.433 & 0.654 & 0.540 & 0.844 & 17.342 & 2.020 & 21.550 &  2.510\\
00/10/26  & 0.224 & 0.020 & 0.196 & 0.291 & 0.874 & 1.301 &  1.771 & 0.246 &  7.890 &  1.095\\
00/11/12  & 1.187 & 0.041 & 0.410 & 0.611 & 0.346 & 0.525 &  8.740 & 0.804 &  7.361 &  0.678\\
00/11/14  & 0.630 & 0.075 & 0.890 & 1.348 & 1.408 & 2.140 & 20.874 & 2.654 & 33.104 &  4.210\\
00/11/17  & 1.120 & 0.029 & 0.747 & 1.117 & 0.668 & 1.001 &  9.486 & 1.241 &  8.487 &  1.110\\
00/11/17  & 0.826 & 0.050 & 0.695 & 1.031 & 0.841 & 1.251 &  8.843 & 1.806 & 10.710 &  2.188\\
01/01/07  & 1.372 & 0.089 & 0.633 & 0.960 & 0.461 & 0.714 & 22.125 & 3.517 & 16.124 &  2.563\\
01/01/19  & 2.630 & 0.256 & 0.792 & 1.182 & 0.301 & 0.554 & 15.580 & 2.970 &  5.930 &  1.130\\
01/02/10  & 2.744 & 0.380 & 0.103 & 0.154 & 0.037 & 3.685 & 68.621 & 8.920 & 25.007 & 3.250\\
01/03/01  & 0.481 & 0.048 & 0.381 & 0.569 & 0.792 & 1.190 & 22.470 & 2.450 & 46.700 &  5.086\\
01/03/23  & 1.766 & 0.063 & 0.577 & 0.859 & 0.326 & 0.498 & 11.900 & 1.660 &  6.740 &  0.941\\
\hline
\textsl{Averages}  & 1.554 & 0.130 & 0.553 & 0.828 & 0.744 & 1.352&
  14.704 & 1.970 & 12.819 &  1.677\\ 
\hline \hline
\end{tabular}
\begin{list}{}{}
\item[] The numbers in columns $2-5$ and $8-9$  are expressed in units of
$10^{30}$ erg/hr. \textsl{Column 1\/}: Date on which the CME occurred; \textsl{Column 2\/}:
Driving power $P_{D}$ associated with a CME; \textsl{Column 3\/}: Error
$\sigma_{D}$ associated with the driving power (\S~2.2); \textsl{Column 4\/}:
Estimate of the magnetic power $P_{M}$ that could be released by the
CME using an estimate of the magnetic field carried by near-Earth magnetic
clouds; \textsl{Column 5\/}: Error $\sigma_{M}$ associated with this estimate
(\S~2.3.1); \textsl{Column 6\/}: Ratio of $P_{M}$ to $P_{D}$;
\textsl{Column 7\/}: Error associated with the quantity $P_{M}/P_{D}$;
\textsl{Column 8\/}: Estimate of the magnetic power
$\widetilde{P_{M}}$ that could be released by the CME using an estimate of the magnetic field
entrained in the CME; \textsl{Column 9\/}: Error $\widetilde{\sigma_{M}}$
associated with this estimate (\S~2.3.2); \textsl{Column 10\/}: Ratio of
$\widetilde{P_{M}}$ to $P_{D}$; \textsl{Column 11\/}: Error associated
with the quantity $\widetilde{P_{M}}/P_{D}$.

\end{list}
\end{table*}

The quantity $\widetilde{P_{M}}$ is an upper limit on the available 
magnetic power, and $\widetilde{\sigma_{M}}$ is the associated error(\S~2.3.1). These are listed in columns 8 and 9 of Table 3. Column 10 of Table 3 gives the ratio of $\widetilde{P_{M}}$
to the required driving power $P_{D}$ and column 11 gives the
error associated with this quantity. The average of the numbers in
column 10 is $12.819 \pm 1.677$. We thus find that the upper limit on
the available magnetic power could be as much as an order of magnitude greater than what is required
to drive the CME. While this discrepancy might seem rather large, 
it may be noted that, 
besides driving the CME, part of the internal magnetic energy
could also be expended in heating the plasma entrained in 
the CME (e.g., Kumar \& Rust 1996) and in overcoming the ``frictional drag''
with the solar wind (e.g., Vrsnak et al. 2004; Cargill 2004).
In-situ measurements of near-Earth magnetic clouds (Burlaga 1988;
Lepping et al 1990; Hu and Sonerup 1998; Mulligan and Russel 2001;
Lynch et al. 2003; Lepping et al 2003) reveal that there
is an appreciable amount of magnetic flux left over after dissipation by
 these means.
 
 The quantity $P_{M}$ is the lower limit on the available magnetic
 power and $\sigma_{M}$ is the associated error(\S~2.3.2). These are
 listed in columns 4 and 5 respectively of Table 3. Column 6 of Table
 3 gives the ratio of $P_{M}$ to the required driving power $P_{D}$
 and column 7 gives the error associated with $P_{M}/P_{D}$. The
 average of the numbers in column 7 is $0.74 \pm 1.35$. We thus find
 that the lower limit on the available magnetic power is an
 appreciable fraction of what is needed to drive a representative CME
 in our sample. The lower limit on the available magnetic power is
 computed on the basis of the magnetic flux detected near the Earth.
 This magnetic flux represents the amount that is left over after
 driving the CME, heating it and overcoming the solar wind frictional
 drag. It is therefore significant that the driving power computed on
 the basis of this residual flux can still account for an appreciable
 fraction of what is needed to drive the CME.

\subsection{Propagation effects and evolution of the white-light flux-rope structure} 

So far, we have been using the generic term ``CME'' to describe the
properties of the flux-rope-like feature that is only a part of the
overall CME phenomenon. It is implicit in our discussion that this
feature comprises a well-defined structure, a system that could
correspond to the flux-rope predicted/invoked in several CME models.
In Paper 1, we suggested that the flux-rope CME propagates as an
isolated system based on our findings of constant total energy for
those events. This result supports the idea that the white-light
signature of a flux-rope CME is indeed a flux-rope.

Perhaps we could get more clues to the nature of the flux-rope
signature by looking into its dynamical evolution. If it is a
flux-rope, we could expect small or no distortion of its shape as it
propagates in the coronagraph field of view. We would also expect
small correlation with the evolution of the other ejecta in the CME.
The evolution of the flux-rope CME can be followed through the
evolution of its center-of-mass. Figure 3 shows the front
and center-of-mass height-time plots for four representative flux-rope
CMEs in our sample. For about half of the events (18/39), the
center-of-mass seems to closely track the evolution of the front. The
events of 1997/04/13 and 1998/05/07 shown in Fig 3 are examples of
such CMEs. 
\begin{figure*}
\centering
\includegraphics[width=17cm]{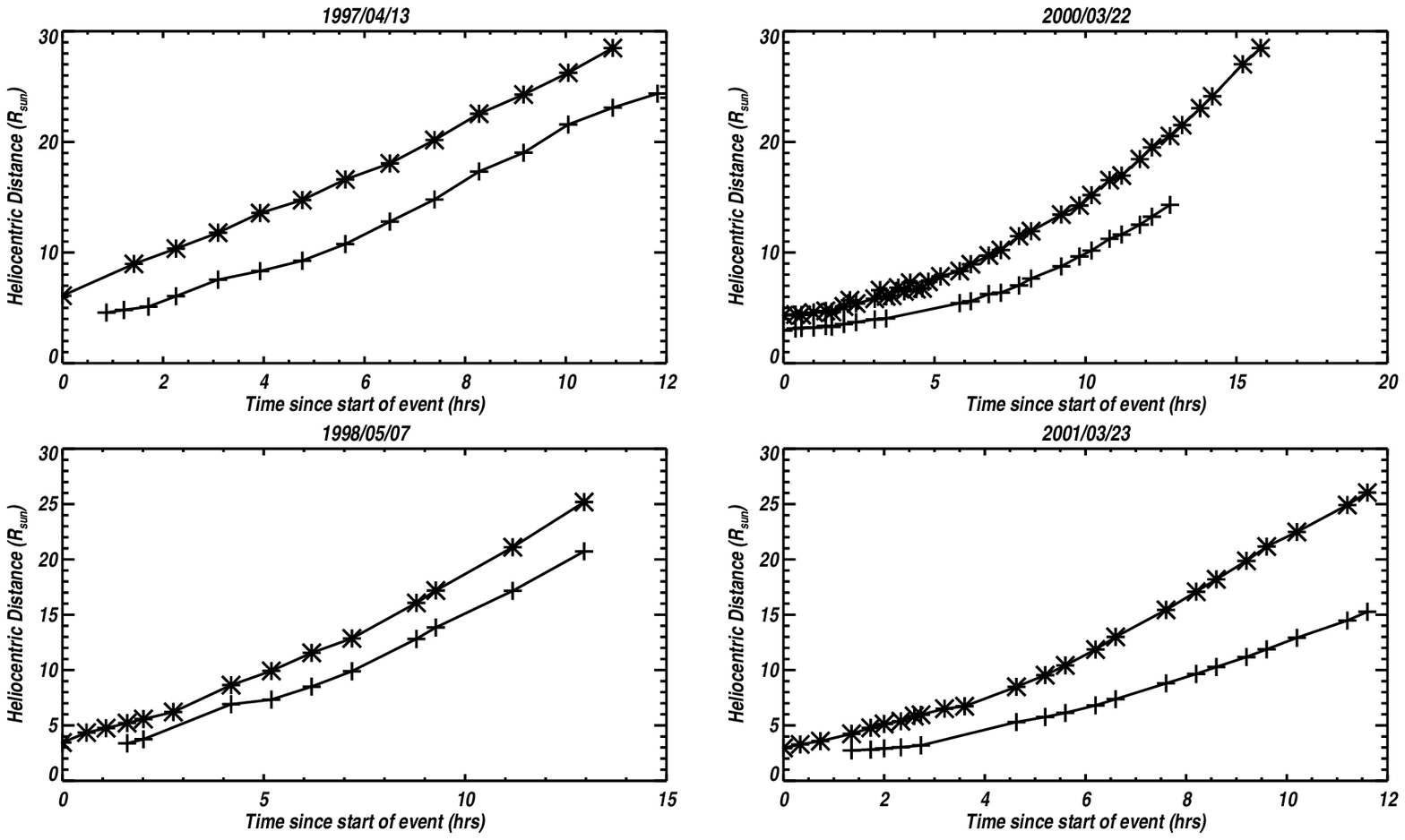}
\caption{Representative front (\textsl{stars}) and center-of-mass
  (\textsl{crosses}) height-time plots for flux-rope CMEs. The event
  date is shown on each plot. The left side panels show events where
  the front and center-of-mass propagate with similar speeds and/or
  accelerations. The right side panels show events where the
  center-of-mass appears to decelerate relative to the CME front. See
  \S~3.4 for further discussion. }\label{cm_plot}
\end{figure*}
For such CMEs, the flux-rope and the CME front propagate
with similar velocities and no distortion of the flux-rope is
observed. This result supports the idea that the white-light feature
is indeed an isolated magnetic structure. 

The events of 2000/03/22 and 2001/03/23 shown in Fig 3 are
representative of the other half of our CME sample (19/39). For these
CMEs, the center-of-mass seems to decelerate relative to the CME front
as is evident from the diverging height-time curves. This is caused by
a progressive center-of-mass shift towards the back of the flux-rope.
The location of the center-of-mass is biased towards the location of
the brightest pixels within the flux-rope structure.  Thus, the shift
of the center-of-mass is due to a brightness increase at the back of
the flux-rope, which is equivalent to mass accumulation at that
location. An inspection of the LASCO mass images supports our
conclusion. It appears that the flux-rope structure of the CME
propagates at a slower speed than the other ejecta coming behind the
main CME structure (the post-CME coronal outflow) which results in the
accumulation of mass at the back of the flux-rope. This is exactly
what one would expect if the CME core is a low beta structure, a
flux-rope, propagating in the solar wind flow. The same behavior has
also been seen in 3D MHD models of erupting fluxropes (Lynch et al
2004). We believe that these observations strongly indicate
that the white light ``flux-rope''-like feature is indeed a
magnetically closed structure; a flux-rope. We also suggest that the
same effect is responsible for the so-called ``disconnection'' or
``V-shaped'' features mentioned often in the literature. In that case,
only the back of the flux-rope is visible either because of the
sensitivity of the instrument or because of the low density of the
white-light flux-rope.

\subsection{Association with prominences}

In theories of filament formation (Karpen et al. 2003 and references
therein), flux-rope structures are commonly associated with either the
filament itself or with large-scale structures within which the
filament lies. Most flux-rope models of CMEs also assume that
prominence material is contained inside the flux-rope. It is therefore
tempting to take the observations of flux-rope-like structures in
white-light coronagraphs as evidence of the existence of flux ropes
in the solar atmosphere, and look for the association of
filament/prominence eruptions with these events. However, the
relationship between pre-existing flux ropes and white-light CMEs is
still unclear from an observational point of view. To see if our
particular sample of CMEs can shed some light on this issue, we
searched for evidence of eruptive prominence/filament associated with
the CMEs we studied. We mainly used the EIT 195\AA\ images because it
is easier to correlate the LASCO/EIT databases. We also used the NOAA
lists of active prominences/filaments, the Nobeyama radioheliograph
database of limb events, and Big Bear H$\alpha$ movies where available.
Our results are shown in column 7 of Table 1 and column 7 of Table 2.
It was generally easy to discern whether a given event involved a
prominence/filament eruption. For the events labeled ``maybe'', we
could see some filament motion or sprays of possibly cool material
(the material appeared dark in the EUV images) but no clear
evidence of large-scale filament/prominence ejection.

We find that 38\% or 15/39 events have a clear association with an
eruptive prominence/filament. A small number of the events (18\% or
7/39) have some indication that chromospheric material was involved,
but we cannot conclusively say whether a large-scale filament was
indeed ejected. Almost half of the events (44\% or 17/39) appear to
have no association with a filament/prominence. This is a somewhat
unexpected result. Given the close morphological resemblance of these
white-light CMEs to flux-ropes, one would expect a closer correlation
between filament eruption and flux-rope-like CMEs. There is always the
possibility that filaments on the far side of the Sun could have been
involved in the events for which we found no filament association on
the visible side or that a filament channel did exist but without
sufficient amounts of cold material to be detected in the images.
Since we do not have any information on the conditions on the far side
of the Sun, we relied solely on the available observations for the
statistics. It might also be possible that these events are associated
with active-region filaments that are generally harder to detect. To
the extent we can tell from our current observations, we conclude
that the flux-rope CMEs in our sample are not strongly correlated with
filament eruptions. Our findings can be contrasted to those of
Subramanian et al. (2001), who found that 59\% of CMEs with signatures
on the solar disk were associated with prominence eruptions.

\subsection{Statistical properties of flux-rope CMEs}

Finally, we can use our relatively large sample of events to derive
statistical properties for the flux rope CMEs. We summarize these
statistics in Fig~4. The distributions of mass and kinetic energies
of the sample are shown in the top panels of Fig~4. 
\begin{figure*}
  \includegraphics[height=7.0in,width=6.5in]{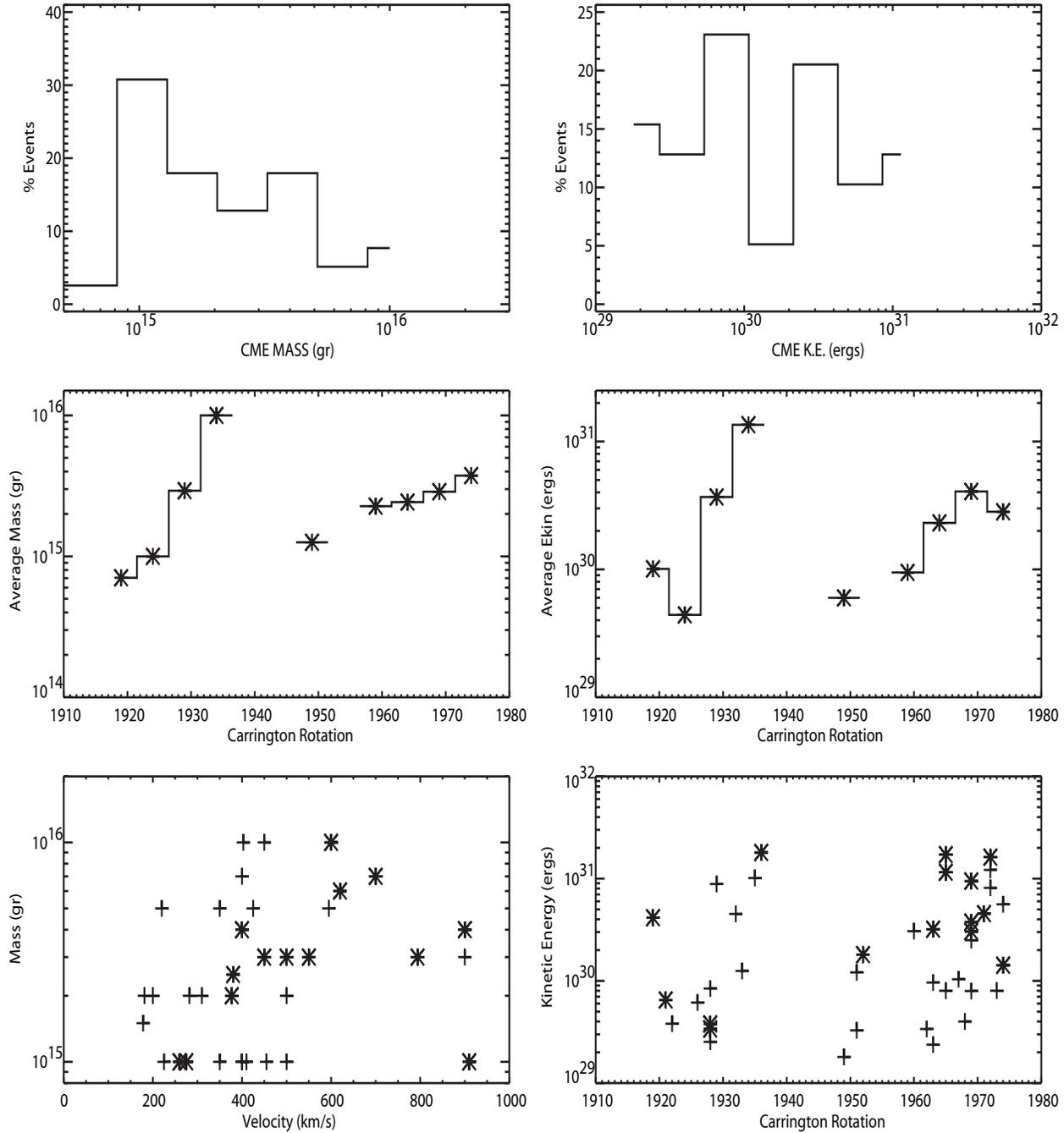}
\caption{Statistics for our flux-rope CME sample. \textsl{Top left:\/}
  Histogram of CME masses. \textsl{Top right:\/} Histogram of CME
  kinetic energies. CME mass as a function of Carrington rotation
  (\textsl{middle left\/}) and CME kinetic energy as a function of
  Carrington rotation(\textsl{middle right\/}). The bins are averages
  over 5 rotations. Scattterplot of CME mass versus front speed
  (\textsl{bottom left\/}) and CME kinetic energy versus Carrington
  rotation (\textsl{bottom right\/}). The stars correspond to the
  events associated with eruptive prominences.}\label{fr_stats}
\end{figure*}

Flux-rope CMEs
have an average mass of $3.1\times10^{15}$ gr and an average kinetic
energy of $4.1\times10^{30}$ ergs. These numbers can be compared to
$1.7\times10^{15}$ gr and $4.3\times10^{30}$ ergs for the average mass
and kinetic energy for the whole sample of LASCO CMEs between
1996-2002 (Vourlidas et al. 2002).  The middle panels of Fig~4 show
the temporal variation of the mass and kinetic energy of flux-rope
CMEs as a function of Carrington rotation. These numbers were
calculated by averaging the measurements over 5 rotations. A sharp
rise in mass and kinetic energy in 1998 (Carrington numbers 1935-1940)
is evident despite the rather small number of events. A similar rise
in the occurrence rate (Gopalswamy et al. 2003) and the average mass
per event (Vourlidas et al. 2002) has been seen in the full sample of
LASCO CMEs. Thus, the rise appears to be a real CME characteristic for
this solar cycle. It is to be noted, however, that LASCO observations
were severely disrupted in the last half of 1998 and early 1999 and
that our statistics have not been corrected for duty cycle. On the
other hand, a slower increase in the flux-rope CME properties since
1999 is also seen in larger CME samples (Vourlidas et al. 2002;
Gopalswamy et al. 2003) and is probably real.  Finally, we look at the
properties of the flux-rope CMEs that are associated with
filaments/prominences. The bottom panels of Fig~4 show the
scatterplots of the mass and kinetic energy of the filament-associated
CMEs (stars) and the rest of the sample (crosses).  It is evident that
filament-associated CMEs are slightly more energetic than the average
CME event. Their average kinetic energy is $6.4\times10^{30}$ ergs,
almost 3 times higher than the average kinetic energy
($2.8\times10^{30}$ ergs) of the total CME sample. The results are
summarized in Table~4.
\begin{table*}
\caption{Statistical properties of flux-rope CMEs}             
\label{table:4}      
\centering 
\renewcommand{\footnoterule}{}  
\begin{tabular}{ccccc}        
\hline \hline
Sample & Average Width &
Average Speed & Average Mass &
Average Kinetic Energy \\
\hline
 & (deg) & (km/s)& ($10^{15}$ gr) & ($10^{30}$ ergs)\\
\hline
Flux-rope CMEs & 90 & 490 & 3.1 & 4.1\\
All LASCO CMEs & 75$^{\mathrm{a}}$ & 417$^{\mathrm{a}}$ &
1.7$^{\mathrm{b}}$ & 4.3$^{\mathrm{b}}$\\ 
\hline \hline
\end{tabular}
\begin{list}{}{}
\item[$^{\mathrm{a}}$] For all CMEs in 1996-2001 (Yashiro et al. 2004)
\item[$^{\mathrm{b}}$] For all CMEs in 1996-2001 (Vourlidas et al. 2002)
\end{list}
\end{table*}

\section{Discussion and conclusions}
We have examined the complete archive of LASCO observations between
1996-2001 and selected the best examples of CMES with a clear flux-rope
structure (39 events). Our measurements suggest that the
``flux-rope''-like structure in the core of these events does indeed
behave as an isolated system, as one would expect from a magnetic
structure (\S~3.4). Overall, we find that only 38\% of these flux rope
CMEs are unambiguously correlated to prominence eruptions (\S~3.3)
which is somewhat suprising given the widely-held notion that the
flux-rope appearance originates from the filament or the cavity above
it. This observation does not preclude the possibility that a filament
channel existed without detectable amounts of prominence material.

We studied the evolution and energetics of the flux rope
structure for these 39 FR CMEs at heights $\sim 2$ R$_{\odot}$--20
R$_{\odot}$.  We find that 69\% of the CMEs in our sample experience a
clear driving power in the LASCO field of view (\S~2.2). We find no
evidence to suggest that these CMEs derive their driving power
primarily via coupling with the solar wind in the range 2--20
R$_{\odot}$.  If this was so, the driving force on the CME would be
directly proportional to its cross sectional area (Eq~4).
However, a scatterplot of driving force on a CME versus its mean cross
sectional area reveals no such trend (Fig~2). On the other hand,
several models for CME propagation rely on different kinds of Lorentz
forces, which ultimately result in the dissipation of its internal
magnetic energy.  To investigate whether the release of the internal
magnetic energy in the CME can possibly provide this driving power, we
adopted two methods. We first used magnetic field measurements obtained
from radio observations of a CME at around 2 R$_{\odot}$. In computing
the available magnetic power using this method, we do not account for
the possible decrease of this magnetic field as the CME propagates
outwards (Eq~2).  It therefore yields an upper limit on the
available magnetic power arising from dissipation of the fields
entrained by the driven CMEs. The upper limit on the available
magnetic power turns out to be an order of magnitude greater than what
is required. We next computed the available magnetic power on the basis
of the flux that is left over in an average near-Earth magnetic cloud
(Eq~5).  Since this is the flux that is left over after
accounting for dissipation in driving the CME from the Sun to the
Earth, heating the CME plasma, and overcoming frictional drag forces,
this method necessarily yields a lower limit on the available magnetic
power. This lower limit is around
$0.74 \pm 1.35$ of what is required to drive the CME. Taken together,
our results thus indicate that the internal magnetic energy of a flux-rope
CME is certainly a viable candidate for propelling it.

\begin{acknowledgements}
SOHO is an international collaboration between NASA
and ESA. LASCO was constructed by a consortium of institutions: the
Naval Research Laboratory (Washington, DC, USA), the
Max-Planck-Institut fur Aeronomie (Katlenburg- Lindau, Germany), the
Laboratoire d'Astronomie Spatiale (Marseille, France), and the
University of Birmingham (Birmingham, UK). We thank Robert Duffin for
helping us identify possible prominence eruptions associated with
the CMEs we studied. We thank the referee for several insightful comments that have
improved the paper.
\end{acknowledgements}


\begin{thebibliography}{}
\bibitem{} Amari, T., Luciani, J. F.; Mikic, Z.; Linker, J.
2000, ApJ, 529, L49
\bibitem{} Antiochos, S. K.; Devore, C. R.; Klimchuk, J. A. 1999, ApJ,
  510, 485
\bibitem{} Bastian, T. S., Pick, M., Kerdraon, A., Maia, D.,
  Vourlidas, A. 2001, ApJ, 558, L65 
\bibitem{} Berdichevsky, D. B., Farrugia, C. J., Thompson, B. J.,
  Lepping, R. P., Reames, D. V., Kaiser, M. L., Steinberg, J. T.,
  Plunkett, S. P., Michels, D. J., 2002, Annales Geophysicale, 20, 891   
\bibitem{} Birn, J., Forbes, T., Schindler, K. 2003, ApJ, 588, 578  
\bibitem{} Brueckner, G.E. et al. 1995, Sol. Phys., 162, 291
\bibitem{} Burlaga, L. F. 1988, J. Geophys. Res., 93, A7, 7217 
\bibitem{} Cargill, P. J., 2004, Sol. Phys., 221, 135
\bibitem{} Chen, J. 1996, J. Geophys. Res., 101, 27499
\bibitem{} Cremades, H., \& Bothmer, V., 2004, A\&A, 422, 307
\bibitem{} Domingo, V., Fleck, B., \& Poland, A. I. 1995, Sol. Phys., 162, 1
\bibitem{} Forbes, T. G. 2000, J. Geophys. Res., 105, 23153
\bibitem{} Gibson, S. E., Low, B. C. 1998, ApJ, 493, 460 
\bibitem{} Gopalswamy, N. et al. 2003, in Proc. of ISCS Symp. on Solar
Variability as an Input to the Earth's Enviroment, Wilson, A. (ed),
ESA SP, in press
\bibitem{} Hu, Q., and Sonnerup. B. U. \"{O}. 1998, Geophys. Res. Lett., 25, 3465
\bibitem{} Karpen, J. T. et al. 2003, ApJ, 593, 1187
\bibitem{} Kliem, B., \& T\"or\"ok, T., 2006, Phys. Rev. Lett., 96, 255002
\bibitem{} Kumar, A., Rust, D. M. 1996, J. Geophys. Res., 101, 15667
\bibitem{} Lepping, R. P., Jones, J. A., and Burlaga, L. F. 1990, J. Geophys. Res.,
  95, A8, 11957 
\bibitem{} Lepping, R. P., Szabo, A., DeForest, C. E., Thompson, B. J. 1997, 
Proc. 31st ESLAB Symp., `Correlated Phenomena at the Sun, in the
Heliosphere and in Geospace, ESTEC, Noordwijk, The Netherlands, 22-25
September 1997 (ESA SP-415, December 1997)
\bibitem{} Lepping, R. P., Berdichevsky, D. B., \& Ferguson,
T. J. 2003, J. Geophys. Res., 108, 1356, 10.1029/2002JA009657
\bibitem{} Lewis, D. J., Simnett, G. M., 2002,
  Mon. Not. R. Astron. Soc., 333, 969
\bibitem{} Lugaz, N., Manchester, W. B., IV, and Gombosi, T. I. 2005, ApJ, 627, 1019
\bibitem{} Lynch, B. J., Zurhuchen, T. H., Fisk, L. A., and Antiochos,
  S. K. 2003, J. Geophys. Res., 108, A6, 1239, 2003 
\bibitem{} Lynch, B. J., Antiochos, S. K., MacNeice, P. J., Zurhuchen, T. H., and Fisk, L. A. Astrophys. J., 617, 589, 2004 
\bibitem{} Manoharan, P. K., Gopalswamy, N., Yashiro, S., Lara, A.,
  Michalek, G., Howard, R. A. 2004, J. Geophys. Res., 109, A06109,
  10.1029/2003JA010300
\bibitem{} Manoharan, P. K., 2006, Solar Physics, 235, 345 
\bibitem{} Mulligan, T., and Russel, C. T. 2001, J. Geophys. Res., 106, 10581, 2001  
\bibitem{} Subramanian, P., Dere. K. P. 2001, ApJ, 561, 372
\bibitem{} Vourlidas, A., Subramanian, P., Dere, K. P., Howard, R. A.,
  2000, ApJ, 534, 456 (Paper 1) 
\bibitem{} Vourlidas, A. et al  2002, in Proc. of the 10th
Europ. Sol. Phys. Meet. 'Solar Variability: From Core to Outer
Frontiers', Prague, Czech Rep., Wilson, A. (ed), ESA SP-506, Dec. 2002, p. 91
\bibitem{} Vourlidas, A. 2004, proceedings of IAU Symposium 226,
  ``Coronal and Stellar Mass Ejections'', Beijing, China, Sept 13 - 17, 2004
\bibitem{}Vrsnak, B., Ru djak, D., Sudar, D., Gopalswamy, N. 2004,
  A\&A, 423, 717 
\bibitem{} Webb, D. F., Cliver, E. W., Crooker, N. U., Cyr, O. C. St.,
  Thompson, B. J., 2000, J. Geophys. Res., 105, 7491
\bibitem{} Yashiro, S. et al. 2004, J. Geophys. Res., 109, A07105, 10.1029/2003JA010282
\bibitem{} Yeh, T., 1995, ApJ, 438, 975
\end{thebibliography}
\end{document}